# Logical difficulty from combining counterfactuals in the GHZ-Bell theorems

**Louis Sica**[1,2]
[1]*Chapman University, Orange, CA & Burtonsville, MD, USA*
[2]*Inspire Institute Inc., Alexandria, VA, USA*
E-mail: lousica@jhu.edu

**ABSTRACT**

In eliminating the fair sampling assumption, the Greenberger, Horne, Zeilinger (GHZ) theorem is believed to confirm Bell's historic conclusion that local hidden variables are inconsistent with the results of quantum mechanics. The GHZ theorem depends on predicting the results of sets of measurements of which only one may be performed. In the present paper, the non-commutative aspects of these unperformed measurements are critically examined. Classical examples and the logic of the GHZ construction are analyzed to demonstrate that combined counterfactual results of non-commuting operations are in general logically inconsistent with performed measurement sequences whose results depend on non-commutation. The Bell theorem is also revisited in the light of this result. It is concluded that negative conclusions regarding local hidden variables do not follow from the GHZ and Bell theorems as historically reasoned.

## 1. INTRODUCTION

The Greenberger, Horn, Zeilinger (GHZ) theorem [1] has achieved a status similar to that of Bell's theorem in its acceptance as a proof that local hidden variables are impossible in quantum mechanics. It has a similarity to Bell's theorem in that it considers a mathematical relation among predicted results of alternative measurements that are not all performed. The alternatives consist of procedures that if performed together are non-commuting, and whose results then depend on their order of execution. Thus, if all the measurements were actually performed, their explicit non-commutation would have to be taken into account in predicting measurement outcomes.

    The use of counterfactuals in no local hidden variables theorems relies on the assumption that counterfactual reasoning is intrinsically sound classically, but not quantum mechanically. However, examples given in Section 2.2 reveal that counterfactual reasoning commonly fails in the classical domain if it neglects the non-commutation of component procedures. It is then shown that parallel reasoning leads to the paradoxical results of the GHZ and Bell theorems in its neglect of non-commutation. The conclusion is that the discrepancy between quantum mechanical eigenvalues and calculations using counterfactuals of non-commuting procedures can

no longer be taken as proof that local hidden variables are inconsistent with quantum mechanical observations.

A definition of counterfactuals and examples showing inconsistencies in their classical use are given in Sections 2.1 and 2.2. In Section 3, the accepted interpretation of the GHZ theorem is described following the treatment by Mermin [2], Home [3], Afriat and Selleri [4], and Greenberger [5], but in a manner showing the roll of counterfactual reasoning. In Section 4, a similar inconsistency due to use of counterfactuals in the Bell theorem is outlined. In this case, the logical inconsistency is manifested by violation of the Bell inequality, an algebraic expression that must be satisfied by cross-correlations of any data sets whatsoever.

## 2. COUNTERFACTUAL REASONING

### 2.1 Definition

In general, the term counterfactual refers to the predicted result of an unperformed act or consequence of a condition that is not true. In the present paper, the definition is further narrowed to distinguish it from various alternatives [6]. If one considers two procedures A and B that do not commute, the result of carrying out a sequence of the two depends on whether A or B is performed first. However, one may consider each procedure in isolation from the other in an exclusive-OR sense. The predicted measurement results of such procedures, that if performed together require non-commutation to be taken into account, are herein termed counterfactuals. (Since measurement outcomes for commuting procedures have simultaneous existence in quantum mechanics, they are of little concern here.)

### 2.2. Flaws in classical counterfactual reasoning

It has been stated in the context of "no-go" theorems for hidden variables that counterfactual reasoning is used frequently in the classical world without any problem: it is logically trustworthy. The author proposes that this belief is in error as will now be shown by classical counter-examples.

We first take note of characteristics of classical non-commuting operations using a semi-facetious example given several years ago by Leon Cohen in a lecture at the Naval Research Laboratory: putting on shoes and socks. Consider this example from the point of view of counterfactuals. One may put on shoes alone, or socks alone in an exclusive-OR sense, and these acts have perfectly well defined meanings. However, one cannot consider these acts in a logical-AND sense unless non-commutation is taken into account. In that case, putting on socks and then shoes gives a different result from putting on shoes and then socks. Thus, converting the logical-OR case to an AND case in the sense of simultaneous existence, or conversion to commutation without conditionality, makes no physical or logical sense.

Classical operations are commonly non-commutative. Consider [7] the rotation of rigid bodies in three-dimensional space. A rotation of $+90^o$ about the x-axis followed by $+90^o$ about the y-axis produces a different final orientation than if these rotations are carried out in reverse order.

Navigation on the surface of the earth is non-commutative: traveling 100 miles north followed by 100 miles west produces a different final position than the same actions carried out in reverse order due to the definitions of north and west on the spherical surface of the earth. However, in the special case where one begins from a position 50 miles south of the equator the results are the same - but then the operations commute.

Consider a beam of polarized light into which are inserted polarizer filters, one at $45°$, one at $90°$ to the initial beam polarization direction. If both polarizers are simultaneously inserted into the beam, the overall transmitted energy depends on their order of placement. If any meaning can be attached to a combination of the two individual states separately, it must be different from that which follows from their actual joint non-commutative placement in the beam. (The measurement is non-commutative because the projection of a vector along a pass axis is transmitted while the perpendicular component is absorbed/discarded.)

Examination of these examples shows that in the classical world, one cannot combine counterfactuals of separate non-commuting operations and expect the result to equal the outcome of the operations performed together. The situation in quantum mechanics is the same on the basis of analysis to be given in Sections 3 and 4. Interestingly, Griffiths has reached a similar conclusion [8] stating that (counterfactual) results of non-commutative operations "cannot be combined to form a meaningful quantum description" in the consistent histories interpretation of quantum mechanics, and that their joint use is meaningless.

## 3. THE GHZ THEOREM

The Pauli spin operators $\sigma_x$, $\sigma_y$, and $\sigma_z$ are used to define three-particle operators $A_1$, $A_2$, $A_3$:

$$A_1 = \sigma_x^1 \sigma_y^2 \sigma_y^3, \ A_2 = \sigma_y^1 \sigma_x^2 \sigma_y^3, \ A_3 = \sigma_y^1 \sigma_y^2 \sigma_x^3, \tag{1}$$

where the superscripts indicate the particle to which the operator is applied. In the theorem, $A_1$, $A_2$, and $A_3$ are ultimately applied to an entangled state of three spin 1/2 particles and each corresponds to a measurement of the product of their spins. Using the anti-commutation properties of the spin operators,

$$\sigma_i \sigma_j = -\sigma_j \sigma_i, \ i \neq j, \ i,j = x, y, z, \text{ and } \sigma_i^2 = 1, \tag{2}$$

and the fact that operators on different particles commute, it is found that $A_1$, $A_2$, and $A_3$ commute. For example, to show that $A_1$ and $A_2$ commute, multiply

$$A_1 A_2 = (\sigma_x^1 \sigma_y^2 \sigma_y^3)(\sigma_y^1 \sigma_x^2 \sigma_y^3) = \sigma_x^1 \sigma_y^1 \sigma_y^2 \sigma_x^2 \sigma_y^3 \sigma_y^3. \tag{3}$$

Using the anti-commutation property of Equation (2),

$$A_1 A_2 = -\sigma_y^1 \sigma_x^1 (-\sigma_x^2 \sigma_y^2) \sigma_y^3 \sigma_y^3 = (\sigma_y^1 \sigma_x^2 \sigma_y^3)(\sigma_x^1 \sigma_y^2 \sigma_y^3) = A_2 A_1. \tag{4}$$

The other commutations may be demonstrated similarly.

One may now consider the product operator $A_1 A_2 A_3$:

$$A_1 A_2 A_3 = (\sigma_x^1 \sigma_y^2 \sigma_y^3)(\sigma_y^1 \sigma_x^2 \sigma_y^3)(\sigma_y^1 \sigma_y^2 \sigma_x^3). \tag{5}$$

Since operations on different particles commute, this may be written

$$A_1 A_2 A_3 = (\sigma_x^1 \sigma_y^1 \sigma_y^1)(\sigma_y^2 \sigma_x^2 \sigma_y^2)(\sigma_y^3 \sigma_y^3 \sigma_x^3), \tag{6}$$

as long as the right to left sequence of operators on each particle is unchanged from Equation (5). Relation (6) may be simplified by using Equation (2), particularly the anti-commutation relation, to obtain

$$A_1 A_2 A_3 = \sigma_x^1 (\sigma_y^2 \sigma_x^2 \sigma_y^2) \sigma_x^3 = \sigma_x^1 \sigma_y^2 (-\sigma_y^2 \sigma_x^2) \sigma_x^3 = -\sigma_x^1 \sigma_x^2 \sigma_x^3. \tag{7a}$$

If one defines $A_4 \equiv \sigma_x^1 \sigma_x^2 \sigma_x^3$, then $A_4 = -A_1 A_2 A_3$, and

$$A_1 A_2 A_3 A_4 = -\sigma_x^1 \sigma_x^2 \sigma_x^3 \sigma_x^1 \sigma_x^2 \sigma_x^3 = -1. \tag{7b}$$

It is emphasized that the minus sign in Equation (7a) that results from multiplying $A_1 A_2 A_3$ is due to the non-commutation of operations on particle 2. Finally from Equation (2) it follows similarly to the examples just given, that $A_1$, $A_2$, $A_3$, and $A_4$ commute for any $|\psi\rangle$.

The GHZ theorem depends on the above state-independent properties of the $A_i$ and further consequences that follow from the fact that they have a common entangled eigenstate

$$|\psi\rangle = \frac{1}{\sqrt{2}}(|\alpha_1\rangle|\alpha_2\rangle|\alpha_3\rangle - |\beta_1\rangle|\beta_2\rangle|\beta_3\rangle). \tag{8}$$

In Equation (8), $|\alpha_i\rangle$ and $|\beta_i\rangle$ for particles $i = 1, 2, 3$ designate the eigenkets of $\sigma_z$ with eigenvalues +1 and -1, respectively. From the well known relations [9]:

$$\sigma_x|\alpha\rangle = |\beta\rangle,\ \sigma_x|\beta\rangle = |\alpha\rangle,\ \sigma_y|\alpha\rangle = i|\beta\rangle,\ \sigma_y|\beta\rangle = -i|\alpha\rangle, \tag{9}$$

and the definitions of $A_1$, $A_2$, and $A_3$, one obtains

$$A_i|\psi\rangle = +1|\psi\rangle,\ i = 1, 2, 3. \tag{10}$$

Then from Equation (7), which follows from the spin anti-commutation relations, action of $A_4$ on $|\psi\rangle$ yields

$$A_4|\psi\rangle = -A_1 A_2 A_3|\psi\rangle = -1|\psi\rangle. \tag{11}$$

Since the $A_i$'s commute and have a common eigenstate $|\psi\rangle$, they are simultaneously measurable on three particles in state $|\psi\rangle$. Thus, for example, the same value of $A_1$ occurs at each of its occurrences in the sequence $A_1 A_2 A_1$. However, a measurement of $A_i$ must be made in such a way that *only the product of the spin values and not their individual values are revealed* [5]. Otherwise, a state produced by measuring an individual $A_i$ would collapse $|\psi\rangle$ to one yielding a specific spin eigenvalue for each particle, and this state would not be an eigenstate of any other $A_i$. Thus, Equations (10) and (11) would no longer hold.

To see this in the case of $A_1$, expand $|\psi\rangle$ in terms of x and y spin basis states:

$$|\alpha_j\rangle = \tfrac{1}{\sqrt{2}}(|\hat{n}_{yj},+1\rangle + |\hat{n}_{yj},-1\rangle),\ |\beta_j\rangle = -\tfrac{i}{\sqrt{2}}(|\hat{n}_{yj},+1\rangle + |\hat{n}_{yj},-1\rangle)$$
$$|\alpha_j\rangle = \tfrac{1}{\sqrt{2}}(|\hat{n}_{xj},+1\rangle - |\hat{n}_{xj},-1\rangle),\ |\beta_j\rangle = \tfrac{1}{\sqrt{2}}(|\hat{n}_{xj},+1\rangle + |\hat{n}_{xj},-1\rangle). \tag{12}$$

This produces a sum of terms, each with spin-product equal to 1:

$$|\psi\rangle = \tfrac{1}{2}(|\hat{n}_{x1},+1\rangle|\hat{n}_{y2},+1\rangle|\hat{n}_{y3},+1\rangle + |\hat{n}_{x1},+1\rangle|\hat{n}_{y2},-1\rangle|\hat{n}_{y3},-1\rangle$$
$$+ |\hat{n}_{x1},-1\rangle|\hat{n}_{y2},+1\rangle|\hat{n}_{y3},-1\rangle + |\hat{n}_{x1},-1\rangle|\hat{n}_{y2},-1\rangle|\hat{n}_{y3},+1\rangle). \tag{13}$$

A measurement of $A_1$ that reveals individual spins would collapse Equation (13) to one of these terms. Suppose it is the first for which all spins are positive. Now consider measurement of $A_2 = \sigma_y^1 \sigma_x^2 \sigma_y^3$: the resulting measured spin product is no longer necessarily +1. On the other hand, if $A_2$ is measured first, its spin product will be the predicted +1, but a following measurement of $A_1$ may now produce a spin product other than +1. The situation is the same for any pair of $A_i$'s if actual spins are revealed. The spin products corresponding to measurement of any one $A_i$ satisfy Equation (10) or (11), but measurement of a second $A_j$, $j \neq i$, no longer necessarily satisfies these relations. *Thus, spin-revealing measurements of different $A_i$ do not commute.*

The usual argument of the GHZ theorem that states that Equations (10) and (11) are inconsistent with local realism, i.e., the existence of hidden variables or pre-existing values for measurement outcomes, is as follows: if local hidden variables supplementing the information in $|\psi\rangle$ are assumed to exist, their values would determine the components of spin found in individual $A_i$ measurements performed on the three particles. If the particles were separated after the formation of $|\psi\rangle$ under the assumption of locality, the measured value obtained for any selected particle spin component would be independent of the *choice* of component measured on any other distant particle. Thus, the value of $\sigma_y$ obtained for particle 3 would be independent of whether one chose to measure $\sigma_y$ or $\sigma_x$ on particle 2. In view of the analysis above, one could measure one of $A_1$, $A_2$, or $A_3$ to obtain

$$m_x^1 m_y^2 m_y^3 = 1, \; m_y^1 m_x^2 m_y^3 = 1, \text{ or } m_y^1 m_y^2 m_x^3 = 1, \qquad (14 \text{ a, b, c})$$

respectively, where the m's, each equal to $\pm 1$, denote numerical values of the measurements. But values of the same symbol occurring in different relations (14 a-c) must be the same based on the condition that the particles do not interact after separation, and on the assumption that the values result from initial conditions determining the measurements of $A_1$, $A_2$ and $A_3$, even though only one of the $A_i$'s can be measured. On the assumption that non-commuting counterfactuals may be combined, the product of (14 a-c) is

$$(m_x^1 m_y^2 m_y^3)(m_y^1 m_x^2 m_y^3)(m_y^1 m_y^2 m_x^3) = m_x^1 m_y^1 m_y^1 (m_y^2 m_x^2 m_y^2) m_y^3 m_y^3 m_x^3 = m_x^1 m_x^2 m_x^3 = 1, \qquad (15)$$

where values for each of the two occurrences of $m_y^i$, $i = 1, 2, 3$ are equal as deduced above.

However, quantum mechanics gives a different result for the combined operation of $A_1, A_2,$ and $A_3$ on $|\psi\rangle$ due to non-commutation of operations on particle 2:

$$A_1 A_2 A_3 |\psi\rangle = -A_4 |\psi\rangle = \sigma_x^1 (\sigma_y^2 \sigma_x^2 \sigma_y^2) \sigma_x^3 |\psi\rangle = -\sigma_x^1 \sigma_x^2 \sigma_x^3 |\psi\rangle = -m_x^1 m_x^2 m_x^3 |\psi\rangle = 1 |\psi\rangle, \quad (16a)$$

so that

$$m_x^1 m_x^2 m_x^3 = -1. \quad (16b)$$

On the assumption that the use of combined counterfactuals in the proof of "no-go" theorems leads to an unavoidable result of classical reasoning, the confirmed measurements of quantum mechanics are concluded to be inconsistent with preexisting or predetermined values for local variables. However, as seen from examples, combined counterfactuals of non-commuting operations do not yield results consistent with those obtained by taking non-commutation into account in the classical realm. Thus, such reasoning does not distinguish between quantum and classical logic. Finally, even if cases exist where a counterfactual construction makes logical sense in and of itself, a different numerical value will be produced than when real experimental operations are performed that are non-commutative.

## 4. BELL'S THEOREM

The present analysis would not be complete without a review of Bell's theorem, and the contribution of the use of counterfactuals of noncommuting operations to its flaws. Previously identified logical problems in the theorem will only be outlined here since their analysis has been given in [10-14].

It is easy to show that the inequality that Bell derived is universally satisfied by cross-correlations of any three or four (as appropriate) data sets consisting of $\pm 1's$ [10,11]. This fact depends *only* on the assumed existence of the data sets, and is independent of any other property such as their origin in random, deterministic, local, or nonlocal processes. Bell did not realize that his inequality resulted from the use of cross-correlation alone in its development. He attributed the result to consciously chosen assumptions: all measurements are represented by a function of random variables (counterfactually) defined at all instrument settings, and locality. He then assumed that the random function he postulated resulted in a second order stationary process [15] in that all correlations could be represented by the same co-sinusoidal function of coordinate differences. Spin measurement operations on a given particle at different instrument settings were mathematically treated as commutative [14].

However, for more than two measurements on a side in Bell experiments, actual performable measurements are noncommutative according to quantum mechanics. Bell mistakenly indicated in his book that this consideration could be ignored by using noncommuting counterfactuals interchangeably with measurements [16] to produce counterfactual experimental data. Of course, since the Bell inequality results from the mere fact of cross-correlation, it is satisfied by the cross-correlations of data sets of commutative second order stationary processes as assumed by Bell and it is thus derivable upon the assumption of such processes, even though it holds generally.

In the quantum mechanical two particle experiments to which this inequality has been applied, consideration of more than one measurement per particle implies that noncommutation must be taken into account. The author has previously identified two experiments yielding more than one measurement per side that could produce data for cross-correlation under this condition. One would use an additional apparatus in tandem on either side of the usual Bell experiment operating in a retrodictive mode [10]. The second would use separate experiments from which correlations conditional on the usual outcomes could be computed [12,13]. Either of these would yield a third correlation that is functionally different from those obtained in standard Bell experiments such that the three would satisfy the Bell inequality as required by basic mathematics.

In practice, the data from Bell experiments have not been cross-correlated, and each pair of correlations is derived from an independent experimental run. If the underlying process were second-order stationary as assumed by Bell, there would only be one correlation functional form to determine, and ensemble averaged cross-correlations would yield the same function as that measured in independent runs. The Bell inequality would be satisfied by this correlation function as measured in independent runs up to small statistical fluctuations. The violation of the Bell inequality by experimentally confirmed cosine correlation functions *proves that the underlying process is not statistically stationary*, contrary to what is widely assumed. *This is consistent with the non-commutative process described by quantum mechanics that predicts different correlation functional forms between some variables.* The logical inconsistencies in the usual interpretation of the Bell theorem begin with combining counterfactuals of noncommutative processes with real data, and are manifested in the violation of the Bell inequality, an inequality that must be universally satisfied by the cross-correlations of any data sets whatsoever consisting of $\pm 1\text{'}s$.

## 5. CONCLUSION

If a theory calculates results of actually performable experiments, the results must logically depend on taking non-commuting operations properly into account. The no-hidden-variables theorems historically contrast quantum results based on non-commutation with classical results based on its neglect. The narrative accompanying these theorems is that neglect of non-commutation of counterfactuals is logically sound in the classical domain, so it is appropriate to attribute the peculiarly inconsistent results that follow to non-locality, or the non-existence of hidden variables or pre-existing values for measurements. But if, as has been shown, the use of counterfactual reasoning in no-hidden-variables theorems is flawed both quantum mechanically *and* classically, the usual paradoxical choices emerging from these theorems no longer have logical motivation. That said, lack of validity of no-hidden-variables theorems does not, in and of itself, imply that local hidden variables exist.

The content of this paper was presented in [17]. The present paper discusses the central idea in greater detail than does the longer [18], while the latter includes additional variations not dealt with here.

## 6. ACKNOWLEDGEMENTS

I would like to thank Mike Steiner of Inspire Institute for useful critical comments on the manuscript, and Armen Gulian and Joe Foreman of the quantum group at Chapman University Burtonsville, MD for many useful discussions relating to the presentation of the material.

### REFERENCES

[1] D. M. Greenberger, M. Horne and A. Zeilinger, "Going beyond Bell's Theorem," In: M. Kafatos, Ed., *Bell's Theorem*, *Quantum Theory and Conceptions of the Universe*, Kluwer, Dordrecht, 2010, pp. 69-72.

[2] N. David Mermin, "Simple Unified Form for the Major No-Hidden-Variables Theorems," *Physical Review Letters*, Vol. 65, No. 27, 1990, pp. 3373-3376. http://dx.doi.org/10.1103/PhysRevLett.65.3373

[3] D. Home, "Conceptual Foundations of Quantum Physics," Plenum Press, New York, 1997, p. 234. http://dx.doi.org/10.1007/978-1-4757-9808-1

[4]  A. Afriat and F. Selleri, "The Einstein, Podolsky, and Rosen Paradox," Plenum Press, New York, 1999, p. 121. http://dx.doi.org/10.1007/978-1-4899-0254-2

[5]  D. M. Greenberger, "GHZ (Greenberger-Horne-Zeilinger) Theorem and GHZ States," In: D. Greenberger, K. Hentschel and F. Weinert, Eds., *Compendium of Quantum Physics*, Springer, Dordrecht, Heidelberg, London, New York, 2009, pp. 258-263.

[6]  L. Vaidman, "Counterfactuals in Quantum Mechanics," In: D. Greenberger, K. Hentschel and F. Weinert, Eds., *Compendium of Quantum Physics*, Springer, Dordrecht, Heidelberg, London, New York, 2009, pp. 132-136. http://dx.doi.org/10.1007/978-3-540-70626-7-40

[7]  H. Goldstein, "Classical Mechanics," Addison-Wesley, Reading, 1980, p. 148.

[8]  R. B. Griffiths, "Consistent Histories," In: D. Greenberger, K. Hentschel and F. Weinert, Eds., *Compendium of Quantum Physics*, Springer, Dordrecht, Heidelberg, London, New York, 2009, pp. 117-122.

[9]  F. Mandl, "Quantum Mechanics," John Wiley and Sons, Chichester, England, 1992, Chapter 5.

[10]  L. Sica, "Bell's Inequalities I: An Explanation for Their Experimental Violation," *Optics Communications*, Vol. 170, No. 1-3, 1999, pp. 55-60. http://dx.doi.org/10.1016/S0030-4018(99)00417-4

[11]  L. Sica, "Bell's Inequalities II: Logical Loophole in Their Interpretation," *Optics Communications*, Vol. 170, No. 1-3, 1999, pp. 61-66. http://dx.doi.org/10.1016/S0030-4018(99)00418-6

[12]  L. Sica, "Correlations for a New Bell's Inequality Experiment," *Foundations of Physics Letters*, Vol. 15, No. 5, 2002, pp. 473-486. http://dx.doi.org/10.1023/A:1023920230595

[13]  L. Sica, "Bell's Inequality Violation Due to Misidentification of Spatially Non-Stationary Random Processes," *Journal of Modern Optics*, Vol. 50, No. 15-17, 2003, pp. 2465-2474.

[14]  J. Malley, "All Quantum Observables in a Hidden-Variable Model Must Commute Simultaneously," *Physical Review A*, Vol. 69, No. 2, 2004, pp. 022118-1-022118-3. http://dx.doi.org/10.1103/PhysRevA.69.022118

[15]  A. Papoulis and S. U. Pillai, "Probability, Random Variables, and Stochastic Processes," McGraw-Hill, New York, 2002, p. 387.

[16]  J. S. Bell, "Speakable and Unspeakable in Quantum Mechanics," Cambridge University Press, Cambridge, 1987, p. 65.

[17]  L. Sica, "Logical Difficulty from Combining Counterfactuals in the GHZ-Bell Theorems," 2013. http://meetings.aps.org/link/BAPS.2013.MAR.R26.10

[18] L. Sica, "Logical inconsistency in combining counterfactual results from non-commutative operations: Deconstructing the GHZ-Bell theorems," arxiv:quant-ph/1202. 0841.